\documentclass[preprint,superscriptaddress,10pt,notitlepage]{revtex4-2}

\usepackage {graphicx}
\usepackage {amssymb}
\usepackage {amsmath}

\newcommand*{\reffig}[1]{Fig.~\ref{#1}}
\newcommand*{\refeq}[1]{Eq.~(\ref{#1})}

\newcommand*{\vect}[1]{\mathrm{\mathbf{#1}}} 

\newcommand*{\erf}{\textrm{erf}\,}

\newcommand*{\imag}{\textrm{Im}\,}

\begin{document}


\title{Femtosecond field-driven on-chip unidirectional electronic currents in nonadiabatic tunnelling regime}



\author{Liping Shi}
\altaffiliation{These two authors contributed equally}
\affiliation{School of Engineering, Westlake University, 18 Shilongshan Road,
  Hangzhou, 310024, China}
\affiliation{Institute of Advanced Technology, Westlake Institute for Advanced Study, 18 Shilongshan Road, Hangzhou, 310024, China}

\author{Ihar Babushkin}
\altaffiliation{These two authors contributed equally}
\email[e-mail: ]{babushkin@iqo.uni-hannover.de}
 \affiliation{Institute of Quantum Optics, Leibniz University Hannover, Welfengarten
1, 30167 Hannover, Germany}
\affiliation{Cluster of Excellence PhoenixD (Photonics, Optics, and
Engineering-Innovation Across Disciplines), 30167 Hannover, Germany}
\affiliation{Max Born Institute, Max-Born-Str.~2a, 12489 Berlin, Germany}

\author{Anton Husakou}
\affiliation{Max Born Institute, Max-Born-Str.~2a, 12489 Berlin, Germany}

\author{Oliver Melchert}
\affiliation{Institute of Quantum Optics, Leibniz University Hannover, Welfengarten
  1, 30167 Hannover, Germany}
\affiliation{Cluster of Excellence PhoenixD (Photonics, Optics, and
  Engineering-Innovation Across Disciplines), 30167 Hannover, Germany}

\author{Bettina Frank}
\affiliation{Universit\"at Stuttgart, Pfaffenwaldring 57,
  70569 Stuttgart, Germany}

\author{Juemin Yi}
\affiliation{Institute of Physics and Center of Interface Science, Carl von
  Ossietzky University Oldenburg, 26129 Oldenburg, Germany}

\author{Gustav Wetzel}
\affiliation{Institute of Electronic Materials and Devices, Leibniz University Hannover, Schneiderberg 32, 30167 Hannover, Germany}

\author{Ayhan Demircan}
\affiliation{Institute of Quantum Optics, Leibniz University Hannover, Welfengarten
  1, 30167 Hannover, Germany}
\affiliation{Cluster of Excellence PhoenixD (Photonics, Optics, and
  Engineering-Innovation Across Disciplines), 30167 Hannover, Germany}

\author{Christoph Lienau}
\affiliation{Institute of Physics and Center of Interface Science, Carl von
  Ossietzky University Oldenburg, 26129 Oldenburg, Germany}

\author{Harald Giessen}
\affiliation{Universit\"at Stuttgart, Pfaffenwaldring 57,
  70569 Stuttgart, Germany}
\author{Misha Ivanov}
\affiliation{Max Born Institute, Max-Born-Str.~2a, 12489 Berlin, Germany}

\author{Uwe Morgner}
\affiliation{Institute of Quantum Optics, Leibniz University Hannover, Welfengarten
  1, 30167 Hannover, Germany}
\affiliation{Cluster of Excellence PhoenixD (Photonics, Optics, and
  Engineering-Innovation Across Disciplines), 30167 Hannover, Germany}

\author{Milutin Kovacev}
\affiliation{Institute of Quantum Optics, Leibniz University Hannover, Welfengarten
  1, 30167 Hannover, Germany}
\affiliation{Cluster of Excellence PhoenixD (Photonics, Optics, and
  Engineering-Innovation Across Disciplines), 30167 Hannover, Germany}

\date{\today}



\begin{abstract}

  Recently, asymmetric plasmonic nanojunctions [Karnetzky et. al., Nature Comm. 2471, 9 (2018)] have shown promise as on-chip electronic devices to convert femtosecond optical pulses to current bursts, with a bandwidth of multi-terahertz scale, although  yet at low temperatures and pressures. Such nanoscale devices are of great interest for novel ultrafast electronics and opto-electronic applications. Here, we operate the device in air and at room temperature, revealing the mechanisms of photoemission from plasmonic nanojunctions, and the fundamental limitations on the speed of optical-to-electronic conversion. Inter-cycle interference of  coherent electronic wavepackets results in a complex energy electron distribution and birth of multiphoton effects. This energy structure, as well as reshaping of the wavepackets during their propagation from one tip to the other, determine the ultrafast dynamics of the current. We show that, up to some level of approximation, the electron flight time is well-determined by the mean ponderomotive velocity in the driving field.

\end{abstract}


\maketitle

\section{Introduction}

As known from atomic physics
\cite{keldysh1965ionization,mevel1993atoms,lindner05-double-slit,babushkin17}, with increase of the incident laser intensity, the ionization dynamics undergoes a transition from a relatively slow multiphoton process to fast sub-cycle bursts, referred to as tunneling ionization. The dimensionless Keldysh parameter $\gamma=2\pi\tau_T/T$ characterizes, relative to the optical cycle duration $T$, a typical time $\tau_T=\sqrt{2m\phi}/|eE|$, required for an electron to leave the atom with the ionization potential $\phi$ in an external field of the amplitude $E$ (here $m$ and $e$ are electron mass and charge). In the multiphoton regime ($\gamma\gg1$), electrons require many optical cycles to be ionized. It is easier to describe such a process in the frequency-domain as an absorption of $n=\phi/\hbar \omega$ photons of energy $\hbar \omega$. But in the tunneling regime ($\gamma\ll1$) electrons escape from the nucleus during a small fraction of an optical cycle, and thus the description in the time-domain is more convenient.
Nevertheless, time- and frequency-domain descriptions represent the same process. In particular,  multiphoton dynamics can be also described in  time-domain, as an interference among the electronic wavepackets
\cite{lindner05-double-slit,zimmerman17-unified-pic-ioniz} created at different optical cycles. This fact was utilized in
the Yudin-Ivanov model \cite{yudin2001nonadiabatic}, where both
multiphoton and tunneling regimes were described in a single formula. In this unified description, the time-domain tunneling picture is used. Yet, the electron ionization dynamics
differ in these two regimes, demonstrating sub-cycle features in the tunneling regime and much slower dynamics in the multiphoton one. 
In the intermediate regime, $\gamma \sim 1$, both fast and slow components appear. 

This tunneling picture arose from atomic physics, and has also been
proven valid for photoemission at surfaces of metallic nanotips
\cite{bormann2010tip,kruger2011-nanotip-tunnnel,yalunin11-photoemission-surfaces,kruger12-metal-nanotip-rev,kruger2012interaction,kruger18-rev-nano}.
Although the electronic wavefunctions inside the metal are not
localized, an approximation of localized wavefunctions still yields
good results when considering ionization from metallic surfaces
\cite{yalunin11-photoemission-surfaces,kruger2011-nanotip-tunnnel,kruger2012interaction,kruger18-rev-nano}.
Strong near-field enhancement near the nanotips triggers electrons
near the Fermi level to tunnel through the surface on a sub-cycle
timescale
\cite{bormann2010tip,kruger2011-nanotip-tunnnel,herink2012field,dombi2013ultrafast,piglosiewicz2014carrier,racz2017measurement,park2012strong,vogelsang2015ultrafast,schertz2012field,robin2016strong,kruger18-rev-nano,dombi20rev},
which is of particular interest in ultrafast time-resolved electron
nanoscopy
\cite{priebe2017attosecond,feist2017ultrafast,zhou2019ultrafast,schoetz2019perspective,kruger18-rev-nano,dombi20rev}.
Interference of the wavepackets tunneled at different cycles gives
rise to pronounced peaks in the electron spectra separated by the
energy of the pump photons
\cite{kruger2011-nanotip-tunnnel,kruger2012interaction,kruger18-rev-nano,dombi20rev}.

Recently, ultrafast electron emission from gold dimer nanoantennas
(nanojunctions) with gap sizes down to the few nanometer scale has
attracted great interest for on-chip petahertz electronics
\cite{schoetz2019perspective,rybka2016sub,ludwig2019sub}. When
embedding such a nanojunction into a closed circuit, unidirectional
electronic optically controllable currents bursts, which vary on the
sub-cycle scale, may arise \cite{rybka2016sub,ludwig2019sub}, if
few-cycle driving pulses with controlled carrier-envelope phase (CEP)
are used. 
Especially interesting in this respect is the recent proposal to create  ultrafast unidirectional currents in  asymmetric nanostructures \cite{karnetzky2018towards}, which enables a DC bias without necessity of CEP control of the driving pulse. 

Since gold nanoantennas cannot withstand strong electric fields due to the limitations of photothermal damage as well as near-field nonthermal ablation
\cite{pfullmann2013bow,putnam2017optical,shi2018resonant,shi2018impact}, the quasistatic tunneling regime ($\gamma\ll 1$) is rather impractical. In contrast, the intermediate regime ($\gamma\sim1$) should be much more attractive. However, in this regime, the nature of photoemission and the corresponding electron dynamics are still poorly understood, despite being  of crucial importance for investigating the limitations on the speed and bandwidth of  on-chip ultrafast electronic devices. 
The pioneering works in Ref.\cite{karnetzky2018towards} left several important questions open. For instance, they suggested that dynamical lowering of the barrier (Schottky effect) can significantly reduce the  scaling of multiphoton photoemission with intensity. However, this assumption has no analogies in other systems in strong optical fields like atoms or molecules. Besides, typical velocities of electrons in the nanogap remain unclarified, although they are crucial to understand the bandwidth limits of the plasmonic electronic devices. 

Here we employ asymmetric plasmonic nanojunctions to produce ultrafast unidirectional currents in a way similar to used in Ref. \cite{karnetzky2018towards}. In contrast to Ref. \cite{karnetzky2018towards}, we operate the device at room temperature and in ambient conditions. Furthermore, we develop a modification of Yudin-Ivanov nonadiabatic tunneling model adapted for gold nanostructures. It allows us to clarify the origin of the photoemission currents. Moreover, we investigate ultrafast dynamics of electronic wavepackets in the nanojunctions, and reconsider fundamental limitations in speed and bandwidth of such optical field-driven electronic devices.  

\begin{figure}
\includegraphics[width=14cm]{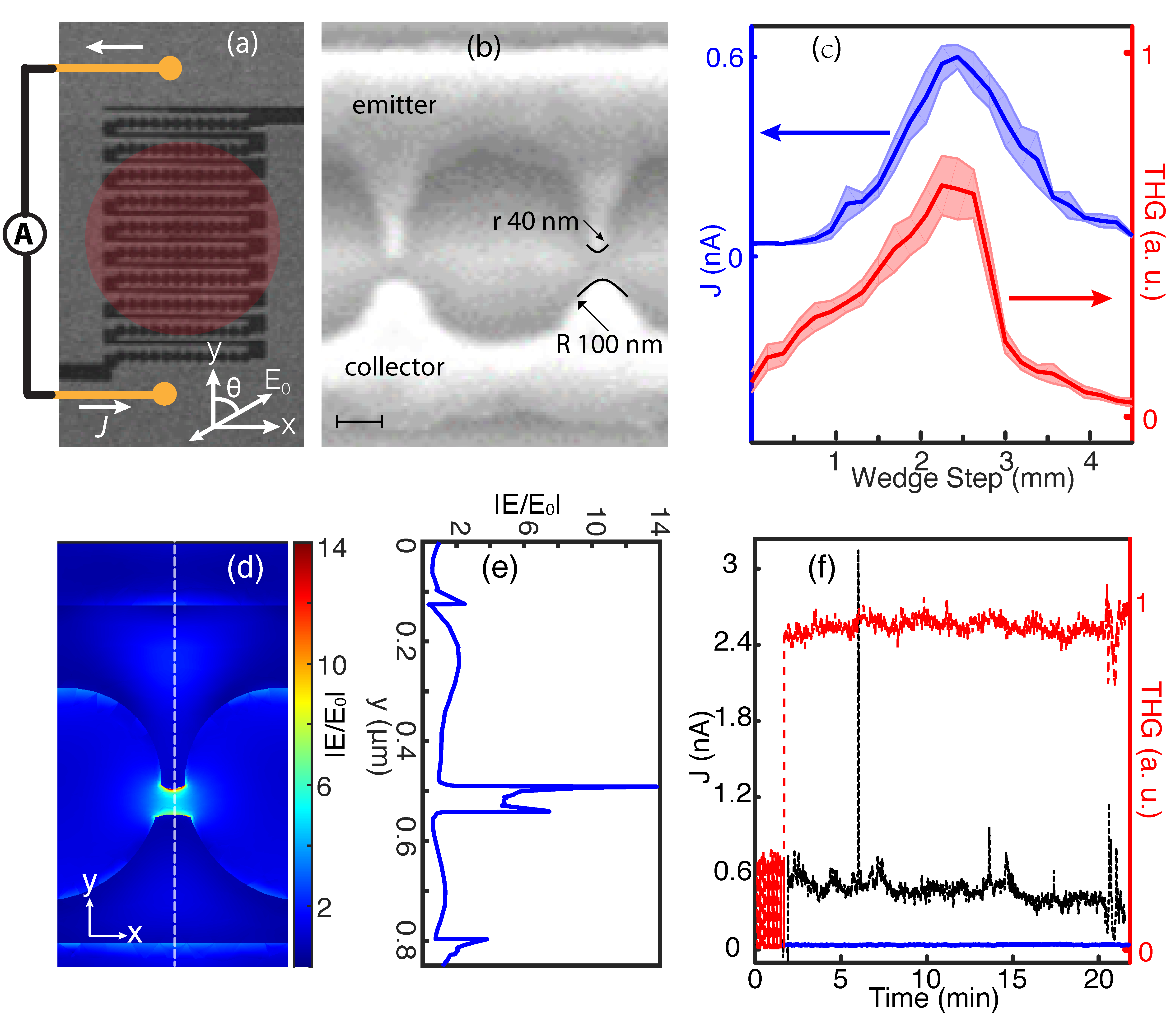}
\caption{\label{fig:1} (a) An overview SEM image of the plasmonic nanodevice. $J$, current; $A$, ammeter; $E_{0}$, optical electric field of incident laser; $\theta$, laser polarization direction with respect to y-axis.
(b) A high-resolution SEM image of two representative nanojunctions, with numerical simulation of electric near-field distribution at the gold-air interface (d). (e) Electric field enhancement factor along the dashed line in (d). (c) Measured photoemission current (blue curve)
and third harmonic signal (red curve) at various thickness of inserted
pair of silica wedges, which controls the laser pulse duration. (f)
Temporal evolution of dark current (blue curve) of the device,
photoemission current (black curve) and third-harmonic generation (red
curve) from the device.  }
\end{figure}

\begin{figure}
\includegraphics[width=12cm]{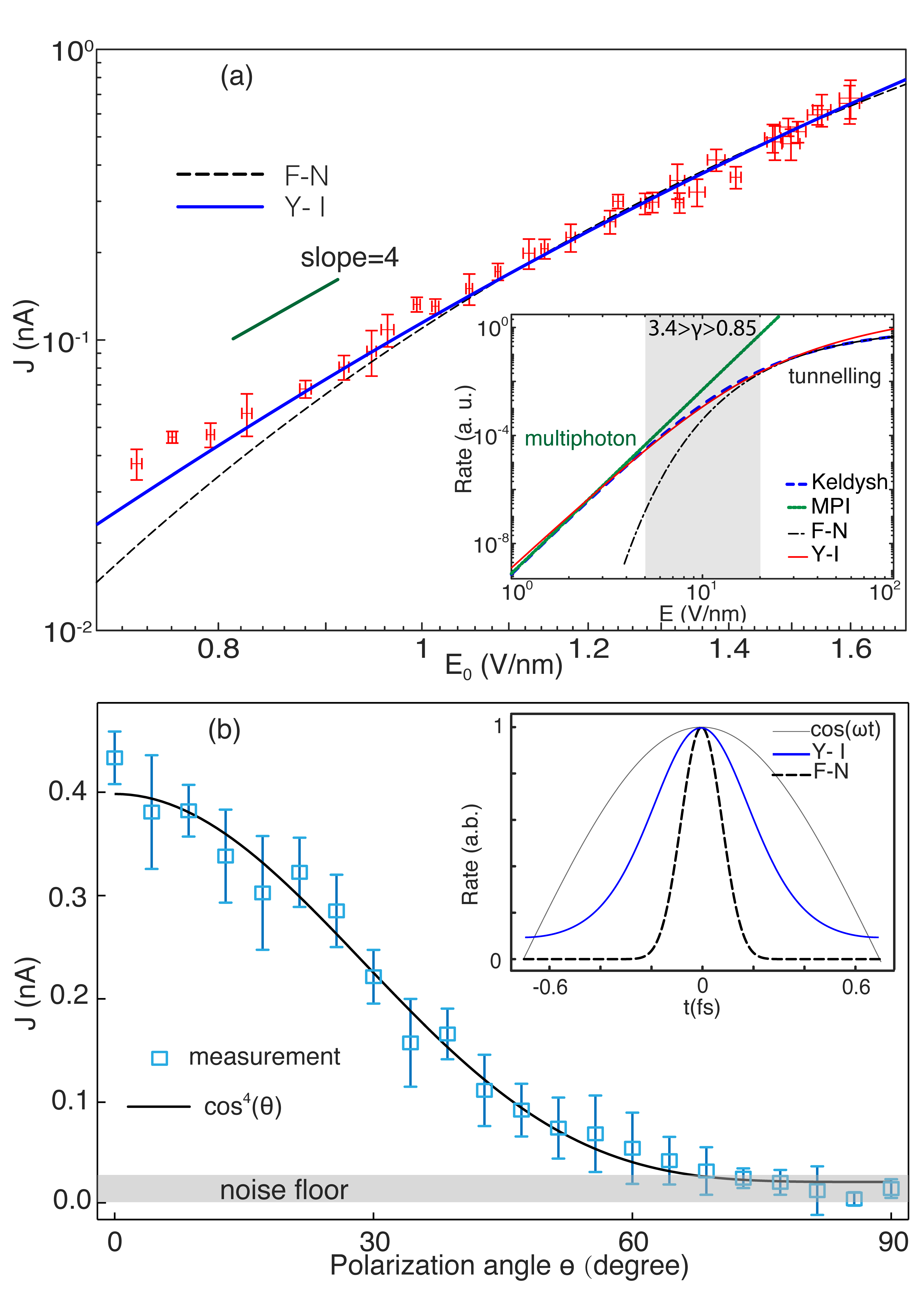}
\caption{\label{fig:2} (a) Experimentally measured photoemission
  current as a function of incident electric field strength $E_{0}$,
  F-N, Fowler-Nordheim model; Y-I, Yudin-Ivanov model. Inset:
  photoemission rate versus laser field strength according to the
  Yudin-Ivanov model (red solid curve) in comparison to multiphoton
  model (MPI, green solid line, rate proportional to $I^4$, where $I$
  is the field intensity), F-N model (black dashed-dotted curve) and
  Keldysh formula (blue dashed curve)\cite{keldysh1965ionization}. (b)
  Photoemission current $J$ (blue squares) versus the laser
  polarization direction $\theta$ in a good agreement with the
  function $J\sim\cos^4\theta$, where $\theta$ is the polarization
  angle (black curve). Inset: Instantaneous photoemission rate over a
  half cycle of laser pulse. Gray curve shows the electric field. Blue
  solid and black dashed curves depict nonadiabatic and quasistatic
  tunneling rates, respectively.
  }
\end{figure}

\section{Experiment}

Femtosecond pulses from a Ti:sapphire oscillator with a repetition rate of $r=100$ MHz are tightly focused onto an array of asymmetric plasmonic nanojunctions. The plasmonic nanojunctions are fabricated by focused ion beam milling of a 100 nm thick Au film on a sapphire substrate. Figure ~\ref{fig:1}(a) depicts an overview scanning electron microscopy (SEM) image of our nanostructure. The laser spectrum spans from 650 to 1000 nm with a central wavelength of 840 nm. Two broadband double-chirped mirrors and a pair of fused silica wedges are employed to control the dispersion of the pulses. We characterize the pulses by dispersion-scan method \cite{shi2019generating} and retrieve the shortest duration to be $\simeq7.6$ fs. The laser beam diameter on the sample is estimated to be around 7 $\mu$m, which corresponds to a simultaneous illumination of about $N\simeq100$ unit cells of the nanojunctions. Therefore, for the highest incident pulse energy of 1 nJ, we estimate the peak electric field of the pump laser to be $E_{0}\sim1.6$ V/nm. Figure ~\ref{fig:1}(b) displays a representative unit cell of the nanojunctions, which consists of a tip-to-tip triangular Au needles with a gap of 50 nm. However, the upper tip is much sharper than the bottom one, leading to an asymmetric distribution of the near-field enhancement, as shown in Fig.~\ref{fig:1}(d) and Fig.~\ref{fig:1}(e). The sharper tip exhibits much higher near-field strength. As a result, an effective negative bias arises from the upper tip to the bottom one, breaking the symmetry of electronic transport. The upper Au needle acts as an emitter electrode and the bottom one functions as a collector. We, therefore, expect to observe a net photoemission current when integrating over the entire pulse width.

We measure the time-integrated photoemission current $J$ by a low-noise amplifier. Meanwhile, the plasmon-enhanced third harmonic generation \cite{hentschel2012quantitative,shi2017self} is employed to optimize the laser dispersion and the spatial overlap between laser focus plane and the plasmonic nanojunctions. 
Figure ~\ref{fig:1}(c) shows the measured photoemission current (blue curve) and third harmonic signal (red curve) at various thickness of the inserted silica wedges. 
The sensitive dependence of the current on the pulse duration confirms that the electron emission is induced by nonthermal processes, because the thermal effects do not depend on the pulse duration for femtosecond laser \cite{plech2006femtosecond}. At the shortest pulse duration, the integrated current reads $J=0.6$ nA, which translates to 40 electrons per pulse in average. As shown in Fig. 1(f), the nanodevice withstands a long-term exposure.

In Fig. 2(a) a log-log dependence of the photoemission current on the incident laser electric field strength ($J-E$ curve) is shown, with an observed slope of $\sim 2$. We also study the integrated current versus the polarization of the incident laser [$J-\theta$ curve, cf. Fig. 2(b)]. Here $\theta$ is defined as the crossing angle between electric field direction and tip-to-tip orientation of the nanojunction. 
The current follows a fourth-order power of $\cos\theta$, corresponding to a $J-E$ curve with a coefficient $n=2$, i.e., $J\propto(E^{2})^{2}=I^{2}$, consistent with directly measured $J-E$ curve in Fig. 2(a). Considering that the electric near-field enhancement factor at a nanotip is inverse proportional to its radius of curvature \cite{novotny2012principles}, as shown by the SEM in Fig. 1 (b), the ratio of photoemission rate from the emitter with respect to the collector is estimated to be $(R/r)^{4}\sim$40. Therefore, we ignore the photoemission from the collector in the below theoretical sections. It should be pointed out that these results also agree with those of Ref. \cite{karnetzky2018towards}. However, we investigate the photoemission process using the time-domain approach rather than the frequency-domain approach, and draw a conclusion opposite to that of Ref. \cite{karnetzky2018towards}, as discussed later.

\section{Theoretical description}

\subsection{Yudin-Ivanov approach}

For the near-field enhancement in our structure the Keldysh parameter is estimated to be in the intermediate range [c.f., gray area in the inset of Fig. 2(a)]. In the framework of atomic physics, a model named Yudin-Ivanov (Y-I) formula \cite{yudin2001nonadiabatic} works well in the range from multiphoton to tunneling regime [red curve in the inset of Fig. 2(a)] and keeps correct inter-and intra-cycle ionization dynamics. Assuming the driving field in the form of $ E(t)= \mathcal E(t) \cos\omega t$, where $\cos\omega t$ denotes the fast oscillating component, and $\mathcal E(t)$ the slow-varying envelope, the cycle-resolved ionization rate $\Gamma$ is given by (in atomic units, that is, frequency $\omega$, time $t$, ionization potential $\phi$, and field $\mathcal E$ are measured in the corresponding Hartree units $\omega_a=0.26$ rad/as, $t_a=24.2$ as, $\phi_a=27.21$ eV, and $\mathcal E_a=514.2$ V/nm):
  \begin{equation}
    \label{eq:YI}
    \Gamma(t)=\frac{\pi}{\tau_{T}}\exp\left(-\sigma_{0}\frac{\mathcal
      E(t)^{2}}{\omega^{3}}\right)\left[\frac{2\kappa^{3}}{\mathcal
      E(t)}\right]^{2Z/\kappa}\exp\left[-\frac{\mathcal E(t)^{2}}{2\omega^{3}}\sigma_{1}\sin^{2}(\omega t)\right].
  \end{equation}
Here $Z$ is the effective atomic charge, $\kappa=\sqrt{2\phi}$,
$\sigma_{0}=\frac12(\gamma^{2}+\frac12)\ln
C-\frac12\gamma\sqrt{1+\gamma^{2}}$,
$C=1+2\gamma\sqrt{1+\gamma^{2}}+2\gamma^{2}$, and
$\sigma_{1}=\ln C$-$2\gamma/\sqrt{1+\gamma^{2}}$. 
The averaged photoemission rate over a single optical cycle  of \refeq{eq:YI} reads (in atomic units):
\begin{equation}
  \label{eq:YI-av}
    \Gamma_{\mathrm{av}}=\frac{\pi}{\tau_{T}}\exp\left(-\sigma_{0}\frac{\mathcal
      E(t)^{2}}{\omega^{3}}\right)\left[\frac{2\kappa^{3}}{\mathcal
      E(t)}\right]^{2Z/\kappa}\left[\frac{2\omega^{3}}{\pi \mathcal E(t)^{2}\sigma_{1}}\right]^{1/2}.
\end{equation}

To proceed further, it is important to understand the basic idea behind the approach leading to \refeq{eq:YI} and \refeq{eq:YI-av}. The population in continuum $W$ is represented as an integral over all partial
amplitudes $a_{\vect p}$ of the ionized electron with the momentum $\vect p$:
$W = \int_{\vect p}|a_{\vect p}|^2d^3p$. The corresponding amplitude $a_{\vect p}$, under reasonable approximations, in particular if the influence of the Coulomb potential after tunneling is neglected, can be obtained as
\begin{equation}
a_{\vect p} \sim \int_{-\infty}^t e^{-iS(t,t')/\hbar}dt',\label{eq:ap}
\end{equation}
where
$S(t,t')$ is the action:
\begin{equation}
  \label{eq:s}
  S(t,t') = (\phi + \frac{p^2}{2m})(t-t') +
  \frac{1}{2m}\int_{t'}^t\left(\vect p + e\vect A(t'') \right)^2dt'',
\end{equation}
 $\vect A(t)$ being the vector potential corresponding to the driving
electric field $E(t)$. Following \refeq{eq:ap}, every
amplitude $a_{\vect p}$ is the result of summation over all partial
amplitudes, each having the phase $S(t,t')$ defined by \refeq{eq:s}.
The first term in \refeq{eq:s} corresponds to phase shift gained by overcoming the barrier, and the second term, so-called Volkov phase, corresponds to the phase electron gains in the electric field. Action $S(t,t')$ changes quickly for all points except the stationary ones, that is, obeying $\partial S/\partial t'=0$. As a result, integration over the fast oscillating argument yields zero everywhere except at   the stationary points (this fact constitutes the essence of  so-called stationary phase, or saddle-point method). The stationary condition results in a complex value of $t'$ equal to $t'_s$, which can be found analytically. The ionization rate can be, up to the insignificant prefactor, calculated as
$\Gamma(t) \sim \exp(-2\imag{[S(t,t'_0)]}/\hbar)$. Under reasonable approximations, this expression can be calculated analytically and gives rise to  \refeq{eq:YI}. 

The Y-I model, as mentioned above, is written for a rather general
case, without the details of the potential. The only parameters
referring to a particular system are the effective charge $Z$ and the
ionization potential $\phi$. For gold nanostructures, we should take
into account the ability of electron density on the surface to
redistribute, on a femtosecond time scale, to ``screen'' the ionized
electron. This screening is in a good approximation described by the
mirror-charge model \cite{jackson62,sahni85}. In this model, if we
consider an ideal metal, the outgoing electron ``feels'' a charge at
the position mirrored relative to the metal surface, but with the
charge sign inverted. The distance from electron to this effective
"ion", represented by mirrored charge, is twice the distance to the
true parent ion which is positioned exactly on the surface. Thus the
attraction force becomes 4 times smaller than for the case of an
electron and a single atom. This attraction force reduction can be
taken into account by introducing the effective charge $1/4$ instead
of $1$ in the Y-I model. Furthermore, in a non-ideal metal, the
effective charge value is modified by a factor
$|(1-\epsilon)/(1+\epsilon)|$, where $\epsilon$ is the complex
susceptibility of the metal. As a result, the effective charge in
\refeq{eq:YI} is governed by the equation:
\begin{equation}
    Z = \frac{1}{4}\left|\frac{1-\epsilon}{1+\epsilon}\right|.
\end{equation}
We fit the experimental data in Fig. 2a by the modified Y-I model and
obtain the field enhancement factor to be $g\approx14.2$. This is a
very reasonable value which is in a good agreement with the numerical
simulation [c.f. Fig. 1(d, e)]. Accordingly, the Keldysh parameter
$\gamma$ in our experiments is evaluated to be in the range from 0.8
to 1.7. The fitting was made without considering the spatial profile
  of the pulse. It is however easy to see that in the present case,
  taking into account the Gaussian profile leads to only a constant
  pre-factor 1/4. Indeed, since we have, with a good precision
  $J\sim E_0^4$, spatial integrating of the current $J(r)$ (were $r$ is
  the radial coordinate) taking into the Gaussian profile of the field
  $E_0\sim \exp(-r^2/2\sigma^2)$ ($\sigma$ is the pulse width) will
  give the constant factor 1/4 in comparison with the same
  integration, performed over the constant-field distribution with the
  same area.

\begin{figure*}
\includegraphics[width=\textwidth]{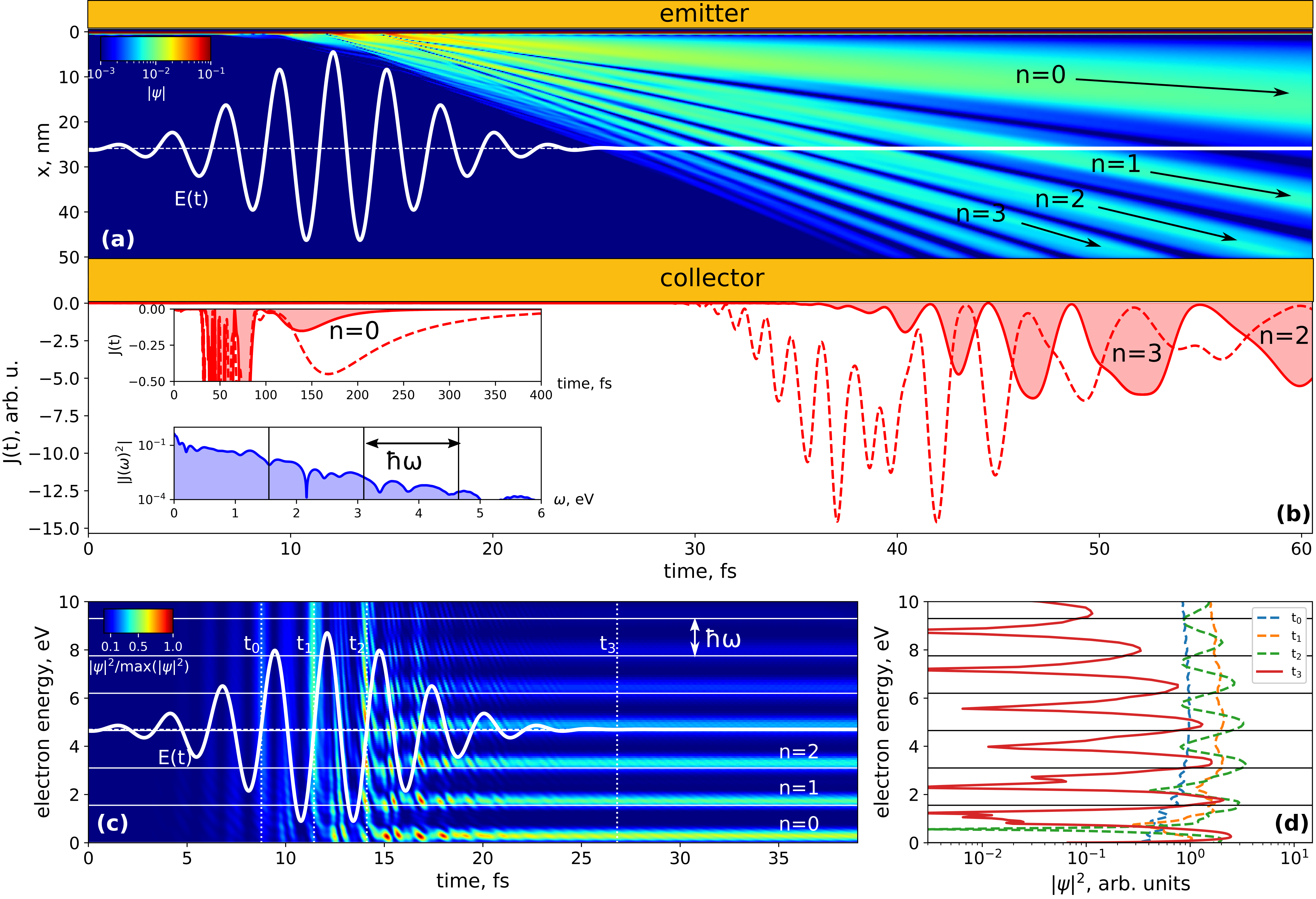}
\caption{\label{fig:3} Dynamics of ionization at the emitter, and
  current at the collector, according to numerical simulations of the
  single-electron problem \refeq{eq:tdse} for $gE_0=12.3$  V/nm and 7
  fs pulse duration. (a) The the modulus of the
  wavefunction of the ionized electron $|\psi(x,t)|$ in dependence on
  $x$ and $t$. White line shows the driving electric field; as the
  electronic wavepacket arises in continuum and propagates, it is
  separated into the well-visible electronic ``beams'' labeled by
  $n=0,1,\ldots$. Each beam [see also (c)] corresponds to an electron
  absorbed approximately $n$ photons from the driving field and thus
  having energy $\approx n\hbar\omega$. (b) The electron current
  $J(t)$ created on the collector surface in dependence on time $t$.
  Dashed red line shows the current created by a homogeneous field
  ($\tilde \alpha=0)$. The upper inset shows $J(t)$ on a larger times
  scale, and the lower inset depicts the spectrum of $J(t)$. The
  vertical lines show the energies corresponding to $n\hbar\omega$.
  (c) Temporal evolution of the free electron energy. White curve
  shows again the driving electric field, whereas the horizontal white
  lines show the energies corresponding to $n\hbar\omega$. (d) Energy
  spectra at the (somewhat arbitrarily selected) times $t_i$ [marked
  in (c)]. In (c) and (d) one clearly sees multiphoton peaks
  $n=1,2,\ldots$ as they arise on the inter-cycle time scale due to
  interference, and become more and more pronounced towards the end of
  the pulse.}
\end{figure*}

As shown in the inset of Fig. 2(a), the Y-I formula approaches the
multiphoton limit at low intensities, and approximates the tunneling
limit as given by the F-N equation at high intensities. We remark that
there is another analytical expression, so-called Keldysh formula
\cite{keldysh1965ionization,zimmermann2019toward}, which works in the
tunneling, multiphoton and intermediate regimes as well. It is
obtained using the same stationary phase method as the Y-I formula,
but the corresponding integrals are taken in the frequency space.
However, the advantages of the Y-I formula are the applicability to
very short (yet single-color) pulses and a possibility to describe the
dynamics inside the laser cycle [c.f. the inset in Fig. 2(b)], whereas
the Keldysh equation was written for the quasi-monochromatic wave and
gives only cycle-averaged ionization rate. For the particular
  parameters presented here there is no possibility to reliably
  differentiate between Y-I and Keldysh formulas. This differentiation
  can be done, for instance, for even shorter pulses in the case if
  the CEP control is implemented. In this situation the change of the
  CEP leads to change of the result in the case of Y-I formula but in
  the framework of the Keldysh formula the result should be
  CEP-independent.

As
depicted in Fig. 2, the current $J$ is nearly proportional to $I^2$.
According to the Y-I model, we identify this $I^2$-law as an
indication that we are in the intermediate regime, and thus no
Schottky effect is necessary to explain this behaviour as it is done
in Refs. \cite{karnetzky2018towards,zimmermann2019toward}. In the
following paragraphs, we return to this point in details.

\subsection{Sr\"odinger equation}

To get deeper insight into the dynamics of the photoemission and subsequent electron propagation between the tips, we simulate the emission by numerical solution of the following one-dimensional
time-dependent Schr\"odinger equation (TDSE) in the Coulomb gauge ($\partial_x A=0$):
\begin{equation}
  \label{eq:tdse}
  i\hbar\frac{\partial \psi(x,t)}{\partial t } =
\frac{1}{2m}\left[ (p + eA(x,t))^2 + V(x)\right]
  \psi(x,t)
\end{equation}
where $\psi(x,t)$ is the electronic wavefunction, $p=-i\hbar\frac{\partial}{\partial x}$, $A(x,t)$ is the vector potential which takes into account spatial field inhomogeneity as a pre-factor $e^{-\tilde \alpha x}$, $\tilde \alpha=1.0$~ns$^{-1}$,  $V(x)$ is a rectangular asymmetric potential 
\begin{equation}
  \label{eq:pot}
V(x)=\begin{cases}
      -V_0,&\text{if $|x|<a$;}\\
      0,&\text{if $x>a$;}\\
      \infty,&\text{if $x<-a$,}
    \end{cases}
\end{equation}
where $a=0.106$ nm, $V_0=16.94$ eV are selected in such a way that i)
the potential has exactly one bound state and ii) the ionization
potential of this bound state equals to the work function of gold (5.1
eV). We note that this potential assumes that the wavefunction inside
the metal is localized. It was shown, however, that this assumption
does not significantly influence the ionization rate
\cite{yalunin11-photoemission-surfaces}. Moreover, the experimentally
observed electron spectra
\cite{kruger2011-nanotip-tunnnel,kruger12-metal-nanotip-rev} and even
electron dynamics \cite{dienstbier21} are good described by this
rather simple approach. Therefore, this approach is widely used to
model the ionization
\cite{hommelhoff2006femtosecond,dombi20rev,kruger18-rev-nano,kruger2011-nanotip-tunnnel,kruger12-metal-nanotip-rev,kruger2012interaction,yalunin2013field,keathley2013strong}.
In the potential defined by \refeq{eq:pot}, ionization can occur only
in the positive direction of $x$. The electrons leaving the emitter
are accelerated by the field and propagate towards the collector.
These electrons are considered to be fully absorbed, that is,
reflection on the potential of the collector is neglected.
We model this by adding to the potential the soft absorbing boundary
$V(x)\to V(x)-i\alpha(x)$ with
$\alpha=\alpha_0\left(1+\erf{\left((x-s)/\delta\right)}\right)/2$,
where $\delta =2.65$ pm, $\alpha_0=1$ au, and $s=50$ nm is the
distance between the emitter and collector. The simulation was made by
a split-step method, with separate evaluation of the terms $\sim p^2$,
$pA + Ap$, $A^2$ and $V$; the action of $p$ was calculated using the
fast Fourier transform.

The resulting dynamics of the electronic wavefunction $\psi(x,t)$ is
shown in Fig. 3(a) for the peak driving field amplitude $gE_0=12.3$ 
V/nm (corresponding to $\gamma=1.68$, clearly in the transient regime)
and pulse duration of 7 fs, according to the experiment. One can see
from Fig. 3(a), that at every positive-field subcycle of the driving
electric field, an ionization event takes place: noticeable part of
the electron is released close to extremum of the electric field. As
one can see from Fig. 3(c), the free electron density increases
starting from the maximum of the electric field, achieves its maximum
one quarter of the cycle later, and then decreases because some part
of electrons returns back. After being ionized, the part of electrons,
which does not return, propagate towards the collector. Interestingly,
upon the propagation the electron wavepacket is separated into
distinct well-visible ``beams'' marked by $n=1,2,\ldots$, every of
them propagating with velocity clearly different from the others. The
field inhomogeneity plays only relatively minor role in this dynamics.
If it is removed [$\tilde \alpha=0$, see dashed line in Fig. 3(b)] the
beam structure remains the same, only most of the electrons move
faster to the collector.

Ionized electrons reach, after some propagation, the collector,
producing the  current $J(t)$ given as
\begin{equation}
  \label{eq:j}
  J(t) = \frac{\hbar}{2im}\left(\psi^*\frac{\partial \psi}{\partial x} - \psi\frac{\partial \psi^*}{\partial x}\right).
\end{equation}
Here, $\psi$ is taken at the surface of the collector. 
The above mentioned ``beams'', which are visible in Fig. 3(a), manifest
themselves as the short spikes of $J(t)$ as can be seen in Fig. 3(b). 

The nature of this dynamics becomes apparent if we consider the
picture in energy space, see Fig. 3(c), where the energy spectrum of
the electrons in dependence on time is shown, as well as in Fig. 3(d),
where energy spectra at specific times $t_i$, $i=0\ldots 3$ are
presented. One can see [cf. for instance $t_0, t_1$] that in the
beginning of the pulse there is no visible structure in the energy
distribution of electrons. Electrons are born with a broad energy
spectrum of more than 10 eV width. The energy structure sets up
gradually during the next few cycles (see the time events $t_2$,
$t_3$), and, finally close to the end of the pulse (the time event
$t_3$) it settles to be peaked around the multiples of the photon
energy $n\hbar \omega$. This very clearly shows that appearance of
multiphoton effects (absorption of $n$ photons) appears only on the
inter-cycle scale, as an interference between newly-born parts of
electron wavepacket and the ones which are already present in the
continuum. These prominent inter-cycle effects in energy space are
well nown in atomic physics
\cite{lindner05-double-slit,zimmerman17-unified-pic-ioniz} but were
demonstrated also for nanostructures, see
\cite{kruger2011-nanotip-tunnnel,kruger2012interaction}.

\begin{figure}
\includegraphics[width=\columnwidth]{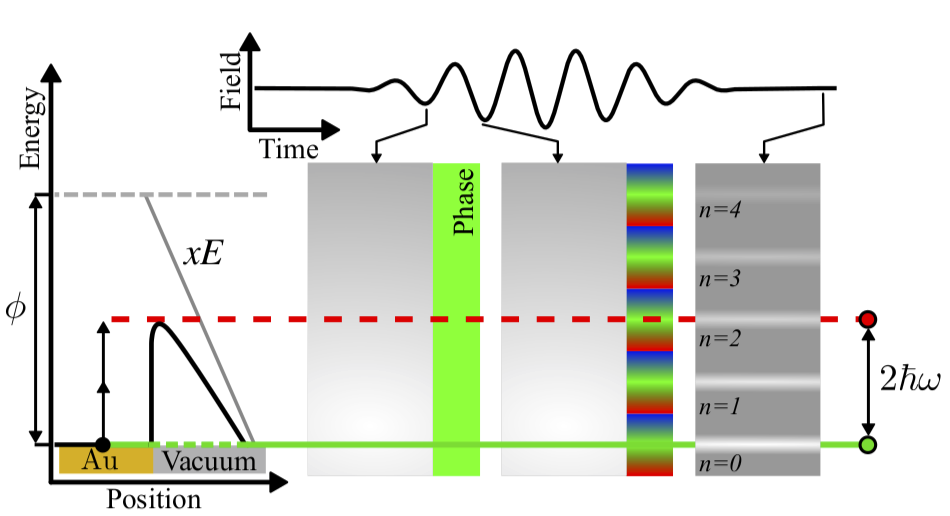}
\caption{\label{fig:4} Schematic representation of different photon  absorption channels by electron ionization. This picture is to be compared to \reffig{fig:3}(c). At every ionization event the electrons are born with a broad unstructured energy spectrum (rectangles with black-white gradient), and phase $\Phi$ (colored gradients) which depends on the temporal position of the ionization event ($\Phi\sim t-t_0$, here $t_0$ is set so that $t-t_0=0$ at the first of two depicted events).  The  interference of these wavepackets leads to visible periodic modulations of the energy distribution after the pulse, which can be interpreted as multiphoton channels $n=0,n=1,\ldots$. Thus, after the end of the pulse one can distinguish between the channels, for instance  direct ionization channel (green horizontal line) and ionization via Schottky effect (red dashed line). The amplitudes of the channels allow to estimate the relative importance of each channel.  }
\end{figure}

In even more clear form this inter-cycle interference is schematically shown in Fig. 4, which illustrates the physics behind the results in Fig. 3(c). As it is presented in Fig. 4, the newborn electrons have a broad spectrum (black-white gradient) and different phases  (color gradient - in the figure we deliberately set the time origin so that the first phase is constant in energy). The interference between parts of the electronic wavepacket created at different cycles leads to clearly visible channels $n=0,n=1,\ldots$ after the end of the pulse. Note that in Fig. 3(c) these dynamics start to set up already in the middle of the pulse since, unless the wavepacket is not the first one [$t_0$ in Fig. 3(c)], the newborn parts of the wavepacket start to interfere with the previously born ones [$t_1,t_2$ in Fig. 3(c)].  

This condition of constructive interference  can be easily obtained as follows: the phase shift $\Phi$ of the electron in the continuum (in the presence of the field) over the optical period $T$ is $\Phi = T(p^2/2m + U_p)/\hbar$, where $U_p=e^2\mathcal E^2/4m\omega^2$ is the ponderomotive energy. At the same time, the electrons in the metal (near the Fermi energy) will experience the relative phase shift $\Phi_F = -T\phi/\hbar$. Thus, the relative phase of two partial wavepackets born in the continuum at the two time instants separated by $T$ is $\Phi+\Phi_F$. The corresponding constructive interference condition $\Phi+\Phi_F=2\pi n$  gives:  $p^2/2m + U_p + I_p = n\hbar\omega$, that is, the subsequent peaks in electron energy $p^2/2m$ are separated by the photon energy $\hbar \omega$. 
This demonstrates that the multiphoton peaks
corresponding to absorption of $n$ photons by an electron appear from the
interference between the electronic wavepackets created by different
optical cycles. Inside the cycle, no such multiphoton effects can be
identified.
One can explain this also in the terms of the Heisenberg uncertainty relation $\Delta p_x\Delta x\ge \hbar/2$: in every ionization event, the electrons are born in the very small region surrounding the surface, of the order of $\Delta x\approx 0.1$ nm, which  means uncertainty in momentum $\Delta p_x$, corresponding to the  kinetic energy in the range from zero to  several eV.
This uncertainty in momentum is partially "regularized" on the longer, intercycle scale as described above, giving rise to the multiphoton energy structure. 

From Fig. 4 one can clearly see how to distinguish the impacts from different ionization channels (accomplished by absorption of different number of photons), and to estimate their relative importance: if a certain channel is present in the ionization process, there must be a corresponding peak in the electron energy distribution after the end of this pulse. The  ``intensity'' of every particular peak allows to estimate the relative impact of different channels. 
To distinguish different channels is important in view of Refs.  \cite{karnetzky2018towards,zimmermann2019toward} where so called Schottky effect is proposed to describe the current-vs-intensity behaviour. 
The Schottky-effect-based explanation suggests that the potential barrier is lowered by the external field (see Fig. 4, red dashed line) so that two instead of four photons are sufficient 
for ionization. This channel corresponds to $n=2$ 
in Fig. 4 and in Fig. 3, in contrast to ``direct'' tunneling which corresponds to $n=0$ (green line in Fig. 4).  

As one can see from Fig. 3(c) and Fig. 4, the peak at $2\hbar\omega$ ($n=2$) which could correspond to Schottky effect indeed appears, but it is also clearly not a dominating one. The peak corresponding to ``direct'' tunneling $n=0$ has even higher amplitude. The ``anomalous'' dependence of the current $J$ on intensity $I$ ($J\sim I^2$) is thus not  explained by the Schottky effect but by the fact that we are in the transition region from tunnel to multiphoton ionization, and the impacts from different channels $n$ have close amplitudes and add up to give the observed  scaling. 

Interestingly, the emergence of the multiphoton energy structure of the electronic wavepacket allows to understand, what happens with an electron as it propagates between the tips [Fig. 3(a)] and thus to interpret the resulting current $J(t)$ in Fig. 3(b). Indeed, different electronic ``beams'' in Fig. 3(a) and different peaks in $J(t)$ in Fig. 3(b) correspond to different peaks in the electron energy in Fig. 3(c). The slowest beam $n=0$ corresponds to the lowest-energy electrons, which absorbed just enough energy to get through the barrier;  $n=1$ corresponds to the electrons which absorbed one photon more, and so on. Every beam corresponds to electrons
which have a mean velocity around  $v=\sqrt{2n\hbar\omega/m}$. Because of diffractive spreading of the wavepacket, its width grows as $\sqrt{t}$. 
The slowest wavepackets spend longer time to overcome the nanogap, and thus spread stronger.  
The resulting structure of the
current $J(t)$ in Fig. 3(b) is thus the collection of peaks, every of them having increasing width due to increasing diffractive spreading. Although the peak with $n=0$ carriers the largest part of the whole electron probability, it is also broadened at strongest, so that as it arrives to the collector, it has a relatively small amplitude (see inset to Fig. 3(b) where this peak is visualized). The same is true for other, not too high values of $n$: the diffractive spreading significantly decreases the corresponding amplitude in $J(t)$. On the other hand, the amount of the electrons contained in the subsequent $n$ decreases as $n$ increases. The global maximum of $J(t)$  is thus the result of this interplay between the diffractive spreading and
energy balance of individual beams. In our configuration, the most intense peaks appear at around $n=7-11$.

As follows from Fig. 3(b), although the shape of the current bunch occurring at the collector is rather complicated, one can define some quantities characterizing it, in particular i) the time delay in respect to the center of the pump pulse, and ii) the duration of the bunch. 
Here we define the above mentioned delay via the position (in time) of the highest peak of $J(t)$, and the width is defined by full-width half-maximum. The defined delay and duration are shown in Fig. 4 for different sizes $s$ of the nanogap. For the particular case of Fig. 3(b) with $s=50$ nm, the current $J(t)$ peaks at around 40 fs (that is, delayed by around 30 fs from the pulse center) and the duration of the current bunch is of the order of 15 fs.

\begin{figure}
\includegraphics[width=\columnwidth]{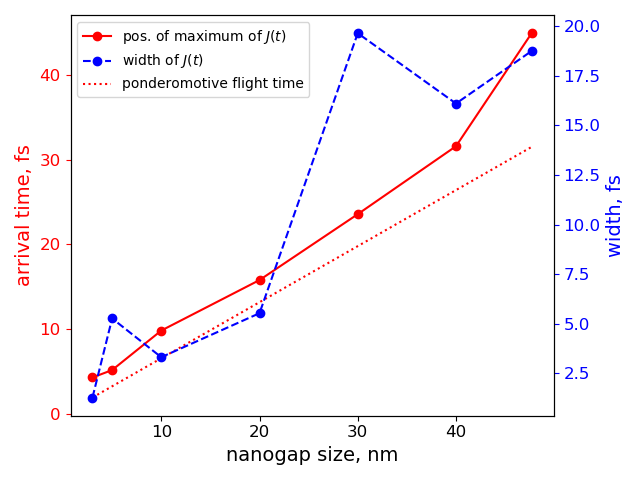}
\caption{\label{fig:5} The position of the maximum of the current $J(t)$ at the collector (red points and red solid line) relative to the pump pulse center (thus denoting the arriving time of the electron at the collector), as well as the width of $J(t)$ (blue points and blue dashed line), as a function of the size of the nanogap, obtained by simulations of \refeq{eq:tdse}. Red dotted line shows the arrival time of an electron at the collector, if the electron's energy is equal to $U_p$, the ponderomotive energy.}
\end{figure}

One can see from Fig. 5 that the delay of the current spike does not,  generally speaking, grow linearly with the gap size. This is explained by the reshaping of the electronic wavepacket described above. This reshaping is, as already mentioned, a result of interplay between the diffractive spreading and electron energy balance. 
Nevertheless, if we compare the delay of $J(t)$ with the na\"{\i}ve estimation for the flight time $t_f=s/v_p$, where $s$ is the size of the nanogap, and $v_p = \sqrt{2U_p/m}$ is the mean velocity corresponding to the ponderomotive energy $U_p = e^2\mathcal E^2/4m\omega^2$ (red dotted line in Fig. 4), we observe quite a good correspondence. This indicates that, although the electron wavepacket has a complicated shape, its propagation can be described with a velocity which corresponds to the mean ponderomotive energy of the electron in the field. 

\section{Conclusion}

In conclusion, we utilized an array of spatially asymmetric
nanojunctions to break the symmetry of the ionization process and to
generate an ultrafast optically switchable on-chip electronic current
at room temperature and under the standard conditions. Generation of
currents up to 0.6 nA by a few-cycle driver pulse with a random
carrier-envelope phase, without using a dc bias, is possible in this
way. Introducing CEP control to this scheme could additionally
  increase the efficiency, but for relatively long pulse durations we
  used here such increase is only minor.  
We have extended the Y-I model, which works well for the tunneling,
multiphoton and intermediate regimes of atoms, to gold nanostructures.
The excellent agreement of the Y-I model allowed to establish the
leading current formation mechanism: the observed data can be well
explained by assuming by nonaddiabatic tunneling through the barrier.
This conclusion is supported by direct simulations of electron
ionization dynamics using the time-dependent Schr\"odinger equation,
which demonstrated that any effects manifesting as an absorption of
several photons occur by the inter-cycle interference of electronic
wavepackets, and are undefined on the sub-cycle time scale. No
signatures of prevailing influence of the Schottky barrier lowering
was found. In contrast, the unusual current scaling $J\sim I^2$ is
explained via the joint influence of all multiphoton channels taking
place in the nonaddiabatic tunneling regime, that is, in the
transition region between tunneling and multiphoton ionization.
Our detailed view of the electron dynamics allowed to determine the
limits on the speed of such devices. We observe that the shape of the
electron wavepacket is rather complicated: the flying electrons are
separated into ``beams'', each of them having the velocity
corresponding to certain number of absorbed photons. Nevertheless, in
average, the flight time of electrons in the nanogap is determined, to
a good precision, by the ponderomotive velocity of electrons in the
driving field. This suggests that the primary way to increase the
speed could be not only to decrease the gap but also to increase the
ponderomotive energy, which does not automatically mean increasing the
peak field: multicolor driving fields could also help at this
\cite{babushkin15multicolor}.









\medskip
\textbf{Acknowledgements} \par 
The authors acknowledge support from Deutsche Forschungsgemeinschaft (DFG) (KO 3798/4-1, BA 4156/4-2, MO 850-19/2, MO 850-23/1) and from German Research Foundation under Germany's Excellence Strategy EXC-2123 and Germany's Excellence Strategy within the Cluster of Excellence PhoenixD (EXC 2122, Project ID 390833453), Lower Saxony through 'Quanten und Nanometrologie' (QUANOMET, Project Nanophotonik). H. G. and B. F. acknowledge funding by ERC (ComplexPlas and 3D Printedoptics) and DFG (SPP1839).
A.H. acknowledges funding from MSCA RISE project ID 823897. C. L. gratefully fully acknowledges the DFG (SPP 1839 and SPP1840) for financial support.


%

\end{document}